\documentclass{aa}
\usepackage{graphicx}

\begin{document}
   \title{Variation of cosmic ray injection across
     supernova shocks}

   \subtitle{}

   \author{H.J.V\"olk
          \inst{1}
          \and
           E.G.Berezhko
          \inst{2}
          \and
          L.T.Ksenofontov
          \inst{2,3}}

   \offprints{H.J.V\"olk}

   \institute{Max-Planck-Institut f\"ur Kernphysik,
                Postfach 103980, D-69029 Heidelberg, Germany\\
             \email{Heinrich.Voelk@mpi-hd.mpg.de}
         \and
              Institute of Cosmophysical Research and Aeronomy,
                     31 Lenin Ave., 677891 Yakutsk, Russia\\
              \email{berezhko@ikfia.ysn.ru}
          \and
              Institute for Cosmic Ray Research, University of Tokyo,
                    Kashiwa, Chiba 277-8582, Japan\\
              \email{ksenofon@icrr.u-tokyo.ac.jp}
             }

   \date{Received month day, year; accepted month day, year}
   
   \authorrunning{V\"olk et al.}
   \titlerunning{Variation of CR injection across SN shock}

\abstract{The injection rate of suprathermal protons into the
diffusive shock acceleration process should vary strongly over the
surface of  supernova remnant shocks. These variations and the
absolute value of  the injection rate are investigated. In the simplest
case, like for SN~1006, the shock can be approximated as being
spherical in a uniform  large-scale magnetic field. The injection rate
depends strongly on the shock obliquity and diminishes as the angle 
between the ambient field and the shock normal increases.  Therefore
efficient particle injection, which leads to conversion of a
significant fraction of the kinetic energy at a shock surface element,
arises only  in relatively small regions near the "poles", reducing
the  overall CR production. The sizes of these regions depend strongly
on the random background field and the Alfv\'en wave turbulence
generated due to the CR streaming instability. For the cases of SN~1006 and
Tycho's SNR they correspond  to about 20, and for Cas~A to between 10
and 20 percent  of the entire shock surface. In first approximation,
the CR production rate,  calculated under the assumption of spherical
symmetry,  has therefore to be renormalized by this factor,  while the
shock as such remains roughly spherical.

   } 
   \maketitle
%

\section{Introduction}

The time-dependent nonlinear kinetic theory of cosmic ray (CR)
acceleration in supernova remnants (SNRs) of Berezhko et al.
(\cite{byk96}) and Berezhko \& V\"olk (\cite{bv97, bv00}), applied to the
remnant of SN~1006, Tycho's supernova, and Cas~A (Berezhko et al.
\cite{bkv02}; V\"olk et al. \cite{vbkr02}; Berezhko et al. \cite{bpv03}),
has demonstrated that the existing data are consistent with very efficient
acceleration of CR nuclei at the SN shock wave, converting a significant
fraction of the initial SNR energy content into CR energy. This energy is
distributed between energetic protons and electrons in a proportion
similar to that of the Galactic CRs.

Recent Chandra observations (Bamba et al. \cite{bamba}) found the fine
structure of the outer shock in SN~1006 to be characterized by extremely
small spatial scales of the X-ray synchrotron emission. These structures
agree very well with the above predictions and provide direct evidence for
the efficient acceleration of nuclear CRs in this object (Berezhko et al.
\cite{bkv03}).

At the same time an essential physical factor which strongly influences
the final CR acceleration efficiency, is contained in our theory as a free
parameter. This is the ion injection rate, i.e. the number of
suprathermal protons injected into the acceleration process per unit area
and unit time. It is described by a dimensionless injection parameter
$\eta$ that is a fixed small fraction of the interstellar medium (ISM)
particles entering the shock front. Assuming spherical symmetry, this
injection implies a source term $Q = Q_{s}\delta (r-R_s)$ due to
(mono-energetic) injection at the subshock in the diffusive transport
equation for the nuclear CR distribution function $f(r,p,t)$
\begin{equation} 
{\partial f\over \partial t}= \nabla(\kappa\nabla f)-\vec w_c \nabla
f+{\nabla \vec w_c \over 3} p {\partial f \over \partial p}+Q,
\label{eq1} 
\end{equation}
where $Q_s$ is written in the form 
\begin{equation} 
Q_s={N_{inj} u_1 \over 4\pi p_{inj}^2} \delta(p-p_{inj}),
\label{eq2} 
\end{equation} 
with
$N_{inj}=\eta N_1$ and $p_{inj}$ denoting the number of injected
suprathermal particles from each unit volume intersecting the shock front
and the momentum of the injected particles, respectively. Here $r$, $t$ and
$p$ denote the radial coordinate, the time, and particle momentum,
respectively; $\kappa$ is the CR diffusion coefficient; $w_c$ is the
radial velocity of the scattering centers, $u=V_s-w$ is the flow velocity
relative to the subshock at $r=R_s$, $V_s=dR_s/dt$ is the subshock
velocity, $N=\rho/m$ is the proton number density, and $m$ denotes the
particle (proton) mass. The subscripts 1(2) correspond to the upstream
(downstream) region.

According to nonlinear acceleration theory, the overall shock transition 
consists of a thin subshock whose thickness is of the order of the gyro 
radius of thermal ions heated in the shock compression, and a CR precursor 
whose much greater spatial extent corresponds to a mean diffusion scale of 
the accelerated particles. Ion injection is thought to occur at the 
subshock. When in the following we losely refer to ``the shock'' in the 
context of injection, then we always mean the subshock.

Unfortunately there is no complete selfconsistent theory of a
collisionless shock transition to the extent that it can predict the value
of the injection rate and its dependence on the shock parameters for all
directions of the shock normal relative to the external magnetic field
vector. For the case of a purely parallel shock (where the shock normal is
parallel to the external magnetic field) hybrid simulations predict quite
a high ion injection rate (e.g. Scholer et al.  \cite{scholer}; Bennett \&
Ellison \cite{benel}) which corresponds to a value $\eta \sim 10^{-2}$ of
our injection parameter. Such a high injection is consistent with
analytical theory (Malkov \& V\"olk \cite{malkv95}, \cite{malkv98}; Malkov
\cite{malkov}) and confirmed by measurements near the Earth's bow shock
(Trattner \& Scholer \cite{trat}).

We note however that in our spherically symmetric model these results can
only be used with some important modification. In reality we deal with the
evolution of the large-scale SN shock which expands into the ISM and its
magnetic field. For example in the case of SN~1006 at the current
evolutionary phase the shock has a size of several parsecs. On such a
scale the unshocked interstellar magnetic field can be considered as
uniform since its random component is characterized by a much larger main
scale of about 100~pc. Then our spherical shock is quasi-parallel in the
polar regions and quasi-perpendicular in the equatorial region. The
leakage of suprathermal particles from the downstream region back upstream
is to leading order dependent upon the shock obliqueness which is
described by the angle between the ambient magnetic field direction and
the shock normal. It is most efficient for a purely parallel subshock and
becomes progressively less efficient when the shock is more and more
oblique (Ellison et al. \cite{elbj}; Malkov \& V\"olk \cite{malkv95}).
Applied to a spherical shock in the uniform external magnetic field it
would mean that only relatively small regions near the poles allow a
sufficiently high ion injection rate which ultimately leads to the
transformation of a significant part (more than a few percent) of the
shock energy into CR energy, whereas the main part of the shock is an
inefficient CR accelerator (V\"olk \cite{vo01}). In this case also the
field amplification due to the CR streaming instability (V\"olk
\cite{vo84}, and in particular Lucek \& Bell \cite{lucb00}, Bell \& Lucek
\cite{belll01}) occurs only near the poles, strongly amplifying the
synchrotron emissivity there.  Such a picture is consistent with the
observed morphology of SN~1006: the most intense synchrotron radiation
comes from two spots (Koyama et al. \cite{koyama};  Allen et al.
\cite{allen}), which we associate with the polar regions.

This very simple picture probably holds only for "ideal" cases like SN 1006
which lies far above the Galactic plane in a very low density and
apparently quite uniform environment. For other objects, presumably
already for Tycho's SNR, and certainly for all core collapse SNRs, the
situation is more complex. Yet the physical arguments which we shall use 
should apply to all of them with appropriate modifications.

In this paper we quantitatively consider the systematic variation of the
ion injection rate across the SN shock surface in a simple
approximation, taking the structure of the ambient magnetic field into
account. For SN explosions into a circumstellar medium that is not
substantially modified by the mass loss from the progenitor star (i.e. SNe
Type Ia, and core collapse SNe whose progenitor stars have zero age main
sequence masses below about 15 $M_{\odot}$), we shall demonstrate that the
size of those shock regions, where efficient injection leading to
efficient CR acceleration is expected to take place, depends strongly on
the random background field and the Alfv\'en wave turbulence generated by
the CR streaming instability. For the case of SN 1006 this corresponds to
about 20\% of the entire shock surface, consistent with the observations.
A similar result is obtained for Tycho's SNR. For the extreme case of
Cas~A, the final phases of stellar evolution -- the red supergiant (RSG)
and the subsequent Wolf-Rayet phase -- have produced a complex
circumstellar pattern of the magnetic field. It is characterized by an
essentially azimuthal mean field of stellar origin with superposed MHD
waves, and strongly modified on large scales by instabilities due to the
radial forces which have their origin in the late appearance of a
high-luminosity Wolf-Rayet phase. An approximate calculation gives the
result that only on a fraction of about 10 to 20 \% of the shock surface
ion acceleration proceeds efficiently.

Electron injection is a rather different problem and our
understanding is much poorer. In many ways, however, electron injection
can be discussed seperately. We will briefly adress it in section 6. While
it is an essential ingredient for the synchrotron and the inverse Compton
emission, to lowest order it neither affects the ion acceleration nor the
overall SNR dynamics.

\section{Spherical shock in a uniform external magnetic field}

\begin{figure}
\centering
\includegraphics[width=7.5cm]{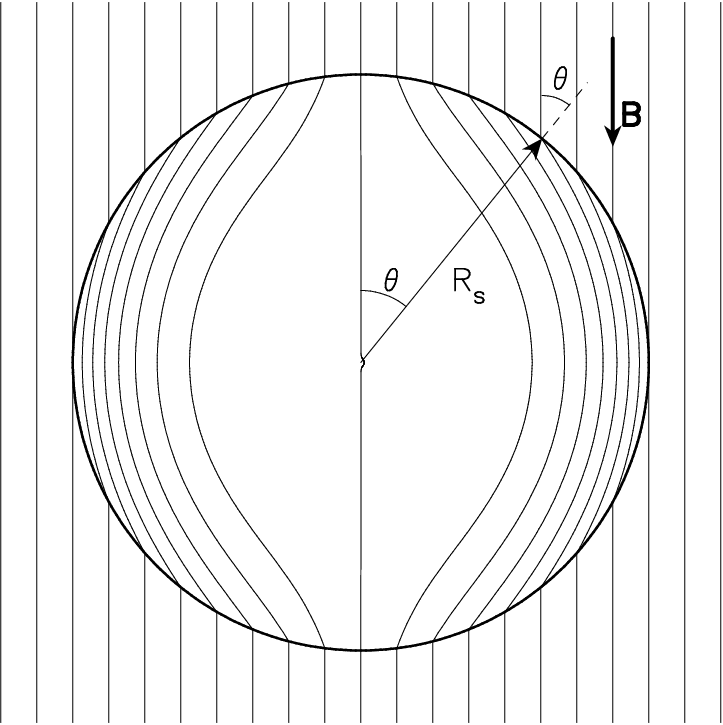}
\caption{Schematic form of the interstellar magnetic field lines, 
modified by a spherically expanding shock ({\it thick circle}).} 
\label{f1}
\end{figure}

Fig.~\ref{f1} schematically illustrates a spherical shock in a uniform ambient
magnetic field $\vec{B}$. In the simplest MHD approximation, the magnetic
field structure in the downstream region is determined by the
compression of the component perpendicular to the shock normal and is
described by the relations:
\begin{equation}
B_{2\parallel}=B_{1\parallel}, 
\hspace{0.5cm}B_{2\perp}=\sigma B_{1\perp},  \label{eq3}
\end{equation}
where $\sigma$ is the shock compression ratio, $B_{\parallel}=B
\cos\theta$, $B_{\perp}=B \sin \theta$.

\begin{figure}
\centering
\includegraphics[width=7.5cm]{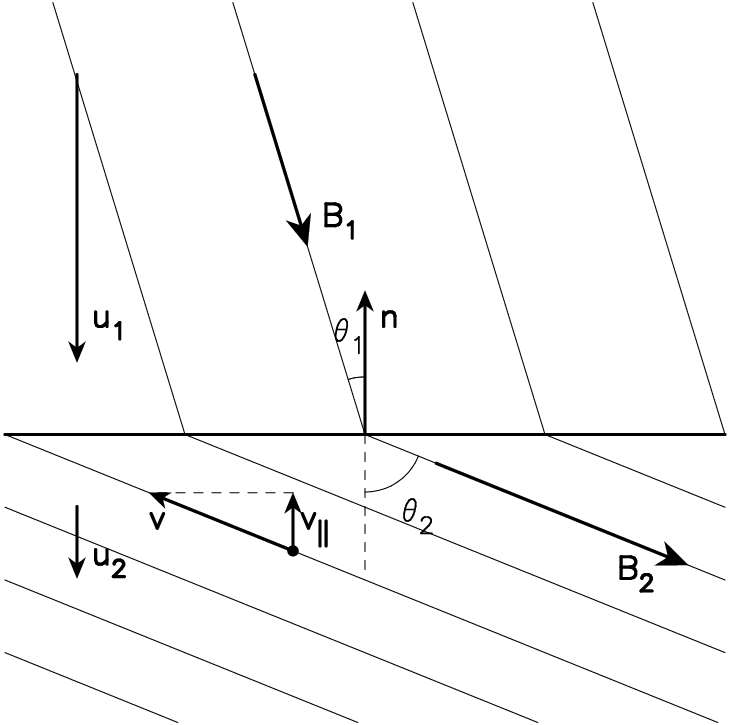}
\caption{Schematic picture  of the local magnetic field and the flow 
structure near
the shock front.} 
\label{f2}
\end{figure}

Since all the scales which characterize the motion of thermal particles
are very much smaller than the shock size, a plane wave approach can be
used for their description. Fig.~\ref{f2} illustrates the magnetic field
structure near the subshock front. The cold upstream plasma is advected
with speed $u_1=V_s$ towards the shock front, is compressed and heated,
and flows with speed $u_2=u_1/\sigma$ into the downstream region. Very
fast particles from the thermalised downstream population, whose velocity
exceeds some critical value $v_{inj}$ and which move towards the shock
front, are able to overtake it and to penetrate into the upstream region.
Since the particle mobility increases with their speed $v$, these
particles can be considered as diffusive and they gain energy in their
random walk across the shock. Asymptotically, for velocities $v \gg
v_{inj}$ (e.g. Malkov \& V\"olk \cite{malkv95}), their pitch angle
distribution approximates an isotropic distribution.

The number of these injected particles is determined by the 
structure of the subshock transition (Malkov \cite{malkov}; see  
Fig.~1 from Malkov \& V\"olk \cite{malkv98}). The most important
physical parameter which determines this number is the particle
velocity parallel to the shock normal $v_{\parallel}$: all particles which
have $v_{\parallel}>v_{inj}$ are considered as injected. It is assumed
that the mean free path of the particles which have a speed higher than
the threshold $v_{inj}$ exceeds the thickness of the subshock.

Assuming the injected
particles to come from the tail of a Maxwellian
distribution we can write
\begin{equation}
\eta_{\parallel}=\exp (-v^2_{inj}/v^2_{T2}),  \label{eq4}
\end{equation}
where $v_{T2}$ is the mean thermal speed of the 
downstream particle population. 
As mentioned above, numerical simulations of parallel collisionless
shocks give an expected injection rate 
$\eta_{\parallel}=\eta(\theta_1=0)\approx10^{-2}$ (e.g. Scholer et al.  
\cite{scholer}; Bennett \& Ellison \cite{benel}), that leads to the
value $v_{inj}=2v_{T2}$ of the injection velocity. 

We assume that the suprathermal particles are strongly magnetized.  
Therefore, for any magnetic field direction, the condition which selects
injected particles is $v_{\parallel}>v_{inj}$. Since
$v_{\parallel}=v\cos\theta_2$ this means that only those particles are
injected which move towards the shock front and have speed
$v>v_{inj}/\cos\theta_2$. Taking into account the relation
\begin{equation}
\cos^2\theta_2=(1+\sigma ^2\tan^2\theta_1)^{-1}  \label{eq5}
\end{equation}
one finds
\begin{equation}
\eta(\theta_1)=\eta_{\parallel}^{1+\sigma^2 \tan^2\theta_1}.  \label{eq6}
\end{equation}

\begin{figure} 
\centering 
\includegraphics[width=7.5cm]{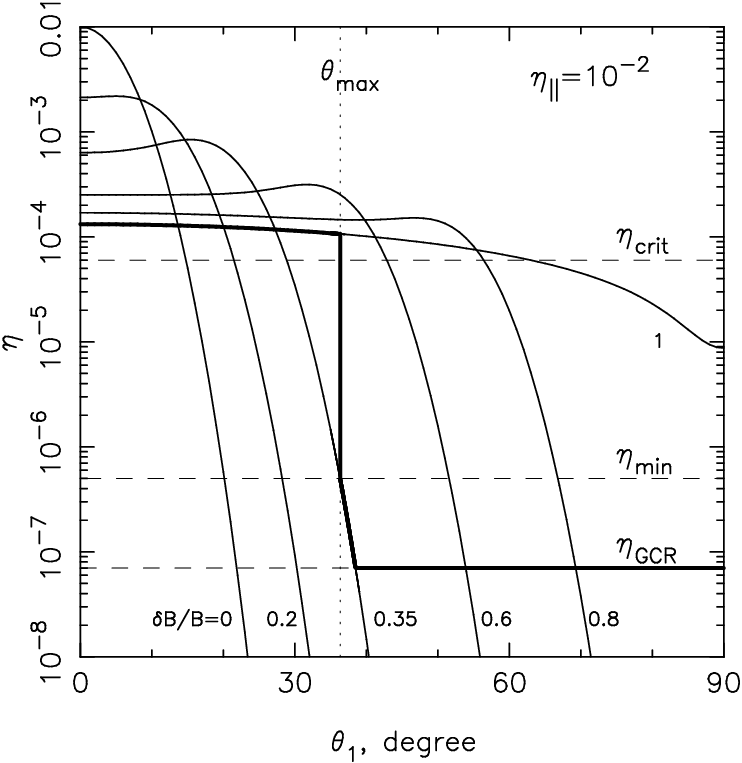}
\caption{The injection rate $\eta$ as a function of the upstream angle
$\theta_1$ between the ambient magnetic field and the shock normal, for
different amplitudes $\delta B/B$ of the upstream random field ({\it solid
lines}), given that $\eta_{\parallel} = 10^{-2}$, for the current
parameters of SN~1006. {\it Dashed lines}
represent the critical injection rate $\eta_{crit}$, the minimal injection
rate $\eta_{min}$ required to provide significant random field
amplification, and the injection rate $\eta_{GCR}$ which is equivalent to
GCR reacceleration. The {\it
thick solid line} represents the expected current injection rate; 
the vertical {\it dotted line} indicates the corresponding
angular region of efficient injection/acceleration.}
\label{f3}
\end{figure}

According to this relation the injection rate goes down quickly with
increasing upstream angle $\theta_1$, as illustrated by the curve which
corresponds to $\delta B=0$ in Fig.~\ref{f3}: for $\theta_1=10^\circ$ the
injection rate is one order of magnitude, and for $\theta_1=20^\circ$ it
is four orders of magnitude smaller than for $\theta_1=0^\circ$.

This simple relation (\ref{eq6}) is a direct consequence of our
assumption that all particles are able to propagate only along the
magnetic field lines. In reality they will to some extent also undergo
cross field motion, either due to drift motions or due to cross field
diffusion. On the other hand this cross field motion can be substantial
only for a relatively high random field component. As shown below, a
high-amplitude random field allows efficient particle injection also
within our simplified approach. 

It is important to note that there exists a so-called critical injection
rate from the solution of the nonlinear kinetic Eq.~(\ref{eq1}), coupled with
the hydrodynamics of the thermal gas.  The value of this critical
injection rate is approximately determined by the expression (Berezhko
\& Ellison \cite{berel}) 
\begin{equation} 
\eta_{crit}=10^{-1}\frac{V_s}{c}
\left( \frac{p_{max}}{mc}\right)^{-1/4},   \label{eq7}
\end{equation} 
where $c$ is the speed of light and $p_{max}$ is the maximum momentum of
the accelerated CRs. It divides the region $\eta>\eta_{crit}$ of the
efficient CR acceleration, when a significant fraction of the shock
energy goes into CR energy, from the region $\eta< \eta_{crit}$ of
inefficient CR acceleration. For $V_s=3000$~km/s and $p_{max}\sim
10^5~mc$ the critical injection rate is $\eta_{crit}=6\times 10^{-5}$.  
This means that in the case of SN~1006, if we do not take into account
the hitherto disregarded random magnetic field component, efficient CR
production is expected to occur only within two polar regions with
$\theta_1<\theta_{max}=14^\circ$. This angular width is considerably
smaller than the one observed (e.g. Koyama et al. \cite{koyama}; 
Allen et al. \cite{allen}).

\section{Magnetic field fluctuations}

The existence of a random magnetic field component $\delta\vec{B}$ on
scales large compared to the thickness of the subshock can change the
value of the injection rate. To study this effect we assume that the
ambient field
\begin{equation}
\vec{B'_1}=\vec{B_1}+\delta\vec{B}   \label{eq8}
\end{equation}
consists of two components, the uniform field $\vec{B_1}$, and a
superimposed, isotropically distributed random component $\delta \vec{B}$.
If the spatial scale of the random component is much smaller than the
shock size $R_s$, one can find the mean injection rate by averaging over
the directions of the random field:
\begin{equation}
\eta=\frac{1}{4\pi}\int d\Omega_{\delta\vec{B}} \eta_{\parallel}^
{1+\sigma^2\tan^2\theta'_1}.  \label{eq9}
\end{equation}
Here $\theta'_1$ is the angle between $\vec{B'_1}$ and the shock normal
direction $\vec{n}$, in analogy to Fig.~\ref{f2}. The averaging procedure
either corresponds to an average over a time interval which is short
compared with the shock age and large compared with the relevant periods
of resonantly scattering field fluctuations for injected particles, or
to a spatial average within the shock region whose size is small
compared with the shock size and large compared with the scales of the
relevant scattering field fluctuations.

The averaged injection rate as a function of angle $\theta_1$ for
different random field amplitudes $\delta B/B$ is shown in Fig.~\ref{f3}. At
small angles $\theta_1$, where the shock is almost purely parallel, field
fluctuations make it more oblique for some fraction of the time.  The
opposite is true for large $\theta_1$. Therefore, as one can see in
Fig.~\ref{f3}, the existence of the random field component leads to a decrease
of the injection rate at small angles and to an increase at large angles
$\theta_1$, so that at the highest values of the turbulent field $\delta
B/B=1$, compared to an ideal parallel shock, 
the injection rate is reduced by almost two orders of magnitude
to a value $\eta(\theta_1)\approx 10^{-4}$ which becomes almost uniform
across an angular range $\theta_1\leq 63^\circ$.

\section{Selfconsistent turbulent field}

The random magnetic field component $\delta \vec{B}$ can be created
selfconsistently by the CR streaming instability in the upstream region
(Bell \cite{bell78}; Blandford \& Ostriker \cite{bla}). The expected
amplitude of the Alfv\'en waves excited due to the CR streaming
instability is determined by the expression (McKenzie \& V\"olk
\cite{mck};  Bell \& Lucek \cite{belll01})
\begin{equation}
\left(
\frac{\delta B}{B}\right)^2=
\frac{V_s}{c_a} \frac{P_{c}}{\rho V_s^2}, \label{eq10}
\end{equation}
where $c_a$ is the Alfv\'en speed,
\begin{equation}
P_c=\frac{4\pi}{3}\int_{p_{inj}}^{p_{max}}dp \, pvf(p) \label{eq11}
\end{equation}
is the CR pressure at the shock front, $p_{inj}$ and $p_{max}$ are the
injection and maximum CR momenta, respectively, and $f$ is the solution
of Eq.~(\ref{eq1}) in which the injection rate appears as a given source term.  
Since we consider effects which can increase an initially very low
injection rate, the shock modification by the CR pressure is
neglected and a plane wave approximation is used. Within this
approach the CR distribution function at the shock front can be written
in the form (e.g. Berezhko et al. \cite{byk96})
\begin{equation}
f=\frac{qN_{inj}}{4\pi p_{inj}^3}
\left(\frac{p}{p_{inj}}\right)^{-q}, \label{eq12}
\end{equation}
where $q=3u_1/(u_1-u_2)$. Taking into account 
that in the case of a strong unmodified shock
$(q=4)$ the appropriate value
of the injected particle speed is $v_{inj}\approx 2V_s$, 
and that relativistic particles with $mc<p<p_{max}$ provide the main
contribution to the CR pressure, we can write
\begin{equation}
\left(\frac{\delta B}{B}\right)^2=
\frac{8c \eta}{3c_a}
\ln\frac{p_{max}}{mc}.  \label{eq13}
\end{equation}
The quantity $\ln (p_{max}/mc)$ depends at most logarithmically on
$(\delta B/B)^2$.  As a consequence it can be seen from this relation
that, once the selfconsistent Alfv\'en wave field $\delta B$ exceeds the
background ISM fluctuation field $\delta B_0$ (see Sect. 7), then any
initially low injection rate leads to the growth of the random magnetic
field in the upstream region which in turn leads to an increase of the
injection rate (see Fig.~\ref{f3}). Equating $(\delta B/B)^2$ to the
background level $(\delta B/B)^2_0$ we can find the minimal initial
injection rate
\begin{equation}
\eta_{min}=\frac{3c_a}{8c\ln (p_{max}/mc)}
\left(\frac{\delta B}{B}\right)^2_0.  \label{eq14}
\end{equation}
On account of the Alfv\'en wave generation by the CR streaming
instability the expected high efficiency injection region is bounded by
the polar angle $\theta_{max}$ determined from the relation
\begin{equation}
\eta(\delta B, \theta_{max})=\eta_{min},  \label{eq15}
\end{equation}
where $\eta(\delta B, \theta_1)$ is the function shown in Fig.~\ref{f3}
for a given random field amplitude $\delta B$. Within the region
$\theta_1<\theta_{max}$ the initial injection rate is high enough so that
accelerated particles are able to increase the level of the turbulence
level which in turn increases the injection rate. This selfconsistent
nonlinear amplification also increases the mean magnetic field strength to an
effective mean field $B$ whose difference to the actual field defines an
effective $\delta B$ (Lucek \& Bell \cite{lucb00}). The process can be
assumed to end when the amplitudes of the Alfv\'en waves become so high
that $\delta B\sim B$ for the effective quantities and their further
growth is prevented by strong nonlinear dissipation processes. Identifying
therefore $(\delta B/B)$ in Fig.~\ref{f3} for $\theta_1<\theta_{max}$ with
the ratio of the effective quantities, the expected selfconsistent
injection rate corresponds to the curve $\eta(\delta B=B,\theta_1)$.

The amplification of the field to an effective field large compared to the
external field $B_1$ is an important aspect of CR production, since it
determines the value of the maximum CR energy. At the same time, this
strongly modified upstream field remains completely randomized. Therefore
we assert that the injection rate within the region
$\theta_1<\theta_{max}$ is not sensitive to the specific value of the
upstream magnetic field $B_1$.

\section{Background cosmic ray acceleration}

Suprathermal particle leakage from downstream is not the only mechanism
that supplies particles to the diffusive shock acceleration process.
Galactic cosmic rays (GCRs) are also subject to further acceleration,
sometimes also called re-acceleration.  We assume that the majority of
GCRs which are participating in this re-acceleration have a momentum
$p\simeq mc$ and therefore their number density can be estimated as
\begin{equation}
N_{GCR}=e_{GCR}/(mc^2),  \label{eq16}
\end{equation}
where $e_{GCR}\approx 0.6$~eV/cm$^3$ is the GCR energy density.  
Comparing the number of CRs with momenta $p\ge mc$, produced in shock
acceleration at a given injection rate $\eta$, with $N_{GCR}$, we can
define a minimal injection rate of suprathermal protons
\begin{equation}
\eta_{GCR}=\frac{cN_{GCR} }{8V_s N_H},  \label{eq17}
\end{equation}
below which GCR re-acceleration becomes more efficient than
acceleration of suprathermal particles.

It can be seen from the above expression that the role of GCR
re-acceleration increases with decreasing shock speed and/or ISM gas
number density. Therefore it can become significant in the diluted ISM
during late SNR evolutionary phases, or for SN explosions in the hot and
low density ISM of elliptical galaxies (Dorfi \& V\"olk \cite{dv96}).

Since re-acceleration of GCRs is almost independent of the polar angle
$\theta_1$ (Drury \cite{dru83}), efficient CR production will occur
practically over the whole SN shock surface, if $\eta_{GCR}$ exceeds the
critical injection rate $\eta_{crit}$.

In the case $\eta_{GCR}< \eta_{crit}$ the role of GCRs can be still
significant if $\eta_{GCR}> \eta_{min}$. Since at large angles
$\theta_1$ GCR reacceleration dominates over the suprathermal particle 
acceleration, Alfv\'en wave generation due to reaccelerated GCRs is also 
higher: they produce in the upstream region an Alfv\'en wave field
\begin{equation}
\left(\frac{\delta B}{B}\right)^2=
\frac{\eta_{GCR}}{\eta_{min}}
\left(\frac{\delta B}{B}\right)^2_0.  \label{eq18}
\end{equation}
Therefore, efficient injection in this case occurs within the 
angular
range $\theta_1<\theta_{max}$ where $\theta_{max}$ is determined from
the relation $\eta(\delta B, \theta_{max})=\eta_{GCR}$, with $\delta B$
from expression (\ref{eq18}).

\section{Electron injection}

The injection of electrons is a different problem and much less well
understood physically. We shall give here a very brief discussion in order
to connect synchrotron and, possibly, inverse Compton emission to the
dynamics and radiation properties of accelerated nuclei in SNRs. 

Suprathermal electrons from the hot downstream region cannot resonantly
scatter on the MHD waves produced by the escaping ions, and other wave
types have to be investigated (Levinson 1996). On the other hand, in more
or less perpendicular shocks reflection of part of the incoming ion
population back into the upstream plasma is possible, exciting
electrostatic waves there in which electrons can be energized. This
energization can certainly inject these electrons into the diffusive shock
acceleration process if the electrons reach GeV energies from which point
on they accelerate like relativistic ions of comparable kinetic energies.  
Models of this kind have been investigated after the pioneering study of
Galeev (1984) by Galeev et al. (1995), McClements et al. (1997) and
Dieckmann et al. (2000). This suggests that, in contrast to ions,
electrons may be even best injected at rather perpendicular shocks. Since
the spatial and temporal scales of this initial electron energization are
very short, this argument can presumably be applied locally to any part of
the shock surface where the shock normal is highly oblique to the local
instantaneous magnetic field. Since furthermore in the quasi-parallel
shock regions the energetically dominant accelerating ions create large
amplitude MHD waves, there may thus be {\it parasitic} electron injection
and acceleration also in these regions.  Altogether, electron injection
might therefore occur more or less everywhere over the SNR shock surface.
Whether this can hold also for subsequent acceleration to the highest
energies, is a quite different matter, and less than clear.

Let us nevertheless assume that electrons are uniformly injected and even
accelerated across the SNR shock. The resulting synchrotron emission is
$\propto B^{(q-1)/2}$ and will therefore still occur predominantly in the
polar regions with their strong turbulent field amplification, despite the
average field compression in the equatorial region (see next section). The
Inverse Compton gamma-ray emission may, however, be fairly uniform over
the SNR surface.

\section{Results and discussion}

As a representative case, we first apply the above formalism to SN~1006.
For the relevant SN~1006 parameters $V_s=3200$~km/s, and
$N_H=0.1$~cm$^{-3}$ we have $\eta_{GCR}=7\times 10^{-8}$.

Since the interstellar random magnetic field is distributed over a wide
range of scales $\lambda$, only part of this spectrum with scales $\lambda
<\lambda _{max}$ has to be considered as the small scale field. The upper
scale can be taken as the diffusive length of the highest energy CRs
accelerated, i.e. $\lambda_{max}= l(p_{max})$. According to our numerical
results (Berezhko et al. 2002), for the current evolutionary phase of
SN~1006, $p_{max}=4\times 10^5~ mc$ and $l(p_{max})=0.08~R_s$. Assuming that
the energy density per unit logarithmic scale interval of the background
ISM turbulent field as function of the spatial length $\lambda$ increases
according to the Kolmogorov law
\begin{equation}
E_w(\lambda)=\left(\frac{\lambda}{L_0}\right)^{2/3}
\left(\frac{B_0}{8\pi}\right)^2,  \label{eq19}
\end{equation}
we find that for scales smaller than $\lambda_{max}=0.08~R_s$ it has a
value $(\delta B/B)^2_0=\int_0^{\lambda_{max}}d\ln\lambda E_w
/(B_0^2/8\pi)=5.3\times 10^{-2}$, or $(\delta B/B)_0=0.23$, taking into
account that the main scale is $L_0=100$~pc. According to
Eq.~(\ref{eq14}), for a typical ISM magnetic field $B_0=3$~$\mu$G and a
moderate maximum CR momentum $p_{max}=10^3mc$, we have in this case
$\eta_{min}=1.7\times 10^{-7}$. Since $\eta_{GCR}<\eta_{min}$, GCR
re-acceleration does not play an important role in this case.

As one can see from Fig.~\ref{f3}, the line $\eta(\theta_1)$ which
corresponds to $\delta B/B=0.23$ intersects the minimal level
$\eta_{min}=1.7\times 10^{-7}$ at $\theta_{max}=31^\circ$.  This means
that an initial injection level represented by this curve $\eta(\theta_1)$
leads within the whole region $\theta_1<31^\circ$ to progressive growth of
Alfv\'en waves, which in turn leads to a corresponding increase of the injection
rate. As argued above, this positive backreaction drives the field
amplification to the point $\delta B/B\sim 1$. Therefore the expected
injection rate $\eta\approx 10^{-4}$ at $\theta_1<31^\circ$ corresponds to
the curve $\eta(\delta B=B,\theta_1)$. Since $\eta >\eta_{crit}$,
efficient CR production will occur in this angular region. For
$\theta_1>\theta_{max}$ the expected injection rate $\eta(\theta_1)$ goes
along the curve $\eta(\delta B/B=0.23, \theta_1)$ and then along
$\eta=\eta_{GCR}$. For all $\theta_1>\theta_{max}$ the expected injection
rate is lower than the critical rate.

Then we have the situation where a sharp boundary $\theta_1=\theta_{max}$
separates the regions of efficient and inefficient CR
injection/acceleration. One has to expect that in reality this boundary is
smoothed as a result of some additional physical process. Possibly the
most important factor is the cross-field diffusion of CRs. Due to the high
level of selfconsistent turbulence within the region
$\theta_1<\theta_{max}$, CR diffusion is almost isotropic.  High energy
CRs with $p>mc$ will therefore be able to penetrate diffusively through
the boundary $\theta_1=\theta_{max}$ in the upstream region. They can also
be accelerated in a finite region $\theta_1>\theta_{max}$. An approximate
CR diffusion length across the regular magnetic field in the upstream
region is their parallel diffusion length $l(p_{max})$. It corresponds to
the angle interval $\Delta \theta_1=(l/R_s)~\mbox{ rad}\approx 5^\circ$.
This means that the smoothed region of efficient CR acceleration extends
up to $\theta_{max}'=\theta_{max}+\Delta \theta \approx 36^\circ$.
Therefore efficient particle injection/acceleration is expected to occur
within a bipolar region of about 20\% of the shock surface. Its size
corresponds rather well to that of the observed bright X-ray synchrotron
emission regions of SN~1006 (e.g. Allen et al. \cite{allen})\footnote{We 
shall not go into speculations here whether the nonlinear wave 
amplification process can spread into SNR surface regions where the shock 
is practically perpendicular, so to say by itself. While such a scenario 
can not be excluded theoretically at this time, we believe that it is 
more fruitful to ask whether the observations give any hint in this 
direction. From present knowledge they do not.}.

\subsection{Renormalization}

According to the above estimate a substantial part of the shock still
efficiently injects and accelerates CRs.  In addition, the overall
conservation equations ensure an approximately spherical character of the
overall dynamics. Therefore, we assume the spherically symmetric approach
for the nonlinear particle acceleration process to be approximately valid
in those shock regions where injection is efficient. To take this
injection fraction
\begin{equation} 
f_\mathrm{re}=1-\cos \theta_{max}'  \label{eq20}
\end{equation}
into account, we need then to introduce a {\it renormalization} factor for
the overall nuclear CR acceleration efficiency, and for all the effects
which it produces in the SNR. According to the above estimate its value in
the case of SN~1006 is $f_\mathrm{re}\approx 0.2$.

\subsection{Synchrotron emission} 

The amplification of the magnetic field in the region of efficient
acceleration leads to an effective downstream field strength of $120\mu$G
strength that exceeds the mean interstellar field of about $3\mu$G by a
factor of 40, according to the comparison of the synchrotron measurements
for SN~1006 with the theoretical model (Berezhko et al. \cite{bkv02}).
Outside the ion injection region, in the ``equatorial'' region, there is
by definition no shock modification and the interstellar field is at most
compressed by a factor of 4, corresponding to a strong adiabatic shock in
the thermal plasma. In fact, the field compression factor is on average
smaller than 4 because the shock is truly perpendicular only on the
equator itself. Therefore the enhancement of the local synchrotron
emissivity in the polar regions relative to the equatorial region exceeds
$10^{1.5}\approx 32$ for the hard particle spectrum $\propto p^{-2}$.  
The spatially integrated emissivities must be adjusted by the ratio
$f_{re}/(1-f_{re})=1/4$ of the polar to the equatorial areas. Therefore
the global excess emissivity of the polar regions is expected to be 8
times larger than the equatorial emissivity on account of the higher
magnetic field.  As a consequence we expect the polar regions to dominate
also the synchrotron radiation, in contrast to the arguments of Ratkiewicz
et al.(\cite{rat}) who assumed that the uniform ISM field is merely
MHD-compressed by the spherical shock, which would then obviously lead to
maximum emissivity in the equatorial region.

\subsection{The case of Tycho's Supernova}

Until now we have given numbers which refer to SN~1006. Also the
discussion of morphological effects was done for this ideal case. Even
though Tycho's SN was also of Type Ia, its present X-ray emission is
dominated by line radiation, and not by nonthermal emission, like SN~1006.  
In addition, the ambient interstellar medium around Tycho appears to be
unexpectedly nonuniform, cf. Reynoso et al. (\cite{reyn97}). As a
consequence, it is not clear whether we can expect a nonthermal X-ray 
morphology as
simple as that of SN~1006. Nevertheless, the magnetic field topology is 
analogous and therefore we shall calculate the renormalization factor in 
an analogous form. 

With a present radius $R_s \approx 2.7$~pc and a mean ambient density $N_H =
0.5$~cm$^{-3}$, Tycho is smaller than SN~1006 and the ambient
density is considerably higher, whereas the present shock velocity $V_s =
3100$~km/s is about the same. This gives $\eta_{min}=2.4\times 10^{-6}
(\delta B/B)^2_0$. Similarly, we have
$(\delta B/B)^2_0 = 3/2(\lambda_{max}/L_0)^{2/3}$ with $\lambda_{max} =
0.27$~pc. Taking the turbulent main scale $L_0$ proportional to the
critical SNR radius at cooling which in turn is proportional to
$N_H^{-1/3}$, we have $L_0=58.5$~pc for Tycho, if $L_0=100$~pc for
SN~1006. This gives $(\delta B/B)_0 = 0.2$, and finally
$f_{re}= 0.18$ for Tycho, roughly similar to the value for SN~1006, as used
in V\"olk et al. (\cite{vbkr02}).

\subsection{Late phases}

Since the energy density of the random component of the background field
$(\delta B/B)^2_0\propto R_s^{2/3}$ increases during the SNR evolution,
the size of the region of efficient injection becomes progressively larger
as well. In the extreme case of a shock with size $R_s\sim 100$~pc --
provided it is still strong enough -- efficient injection/acceleration
occurs across the entire shock surface due to the completely randomized
background field on the scale $R_s$. This situation makes the contribution of
the late SNR evolutionary phases more important, and therefore the
resultant CR energy spectrum produced in such SNRs is expected to be
steeper compared with the model prediction in spherical symmetry.

According to expression (\ref{eq7}) the critical injection rate goes down
proportionally to the shock speed $V_s$, whereas the injection rate
(\ref{eq17}) due to GCRs is inversely proportional to $V_s$. Therefore the
significance of GCR reacceleration in the sense of an injection mechanism
progressively increases during SNR evolution. When the shock speed drops
to the value $V_s \approx 35/\sqrt{N_H/(1~\mbox{cm}^{-3})}$~km/s, $\eta_{GCR}$
exceeds $\eta_{crit}$ and then the GCRs alone lead to efficient CR
acceleration across the whole shock surface. Since the GCR chemical
composition differs from the ISM composition, GCR re-acceleration can play
a role in determining the resultant chemical composition.

\subsection{Renormalization factor for Cas~A}

A much more complicated situation arises in SNRs where the circumstellar
medium is strongly influenced by the wind from the progenitor star. Cas~A
is a prominent example. According to the analysis of the thermal X-ray
emission, and consistent with the observed overall dynamics of Cas~A, the
supernova shock expands into an inhomogeneous circumstellar medium
strongly modified by the intense wind of the progenitor star (Borkowski et
al. \cite{bor96}). It consists of a tenuous inner bubble, created by the
Wolf-Rayet (WR) wind, a dense shell of swept-up, slow red supergiant wind
(RSG) material, and a subsequent free RSG wind.  The ambient
magnetic field structure is not well known in this case. Therefore it is
not possible to perform an equally definite analysis of the expected
injection rate as for Type Ia SNe. As we shall demonstrate below, quite a
reasonable estimate can nevertheless be obtained.

Due to the progenitor star's rotation, the mean large-scale magnetic field
is expected to be almost purely tangential to the shock surface in the
wind material. At the same time, as shown in numerical simulations
(Garcia-Segura et al. \cite{garcia}), the interaction of the fast WR with
the slow, massive RSG wind and the shell formation is accompanied by a
long-wavelength instability. We assume that this instability is able to
strongly randomize the preexisting field in the shell, and to possibly
even amplify it. Therefore the ambient circumstellar magnetic field seen
by the SNR shock in the shell should consist of three components
\begin{equation} 
\vec{B_1}=\langle\vec{B}\rangle +\delta\vec{B_l}+\delta\vec{B_s},
\label{eq21} 
\end{equation} 
a mean field $\langle\vec{B}\rangle$ which is practically everywhere
tangential to the shock surface, a large-scale random component
($\delta\vec{B_l}$), and a small-scale random component
($\delta\vec{B_s}$). The small-scale field component consists of the
copious amount of MHD waves, essentially Alfv\'en waves, emitted by the
central RSG, and is expected to follow a nonlinear cascade towards short
scales, taken here to be of the Kolmogorov type.

\begin{figure} 
\centering 
\includegraphics[width=7.5cm]{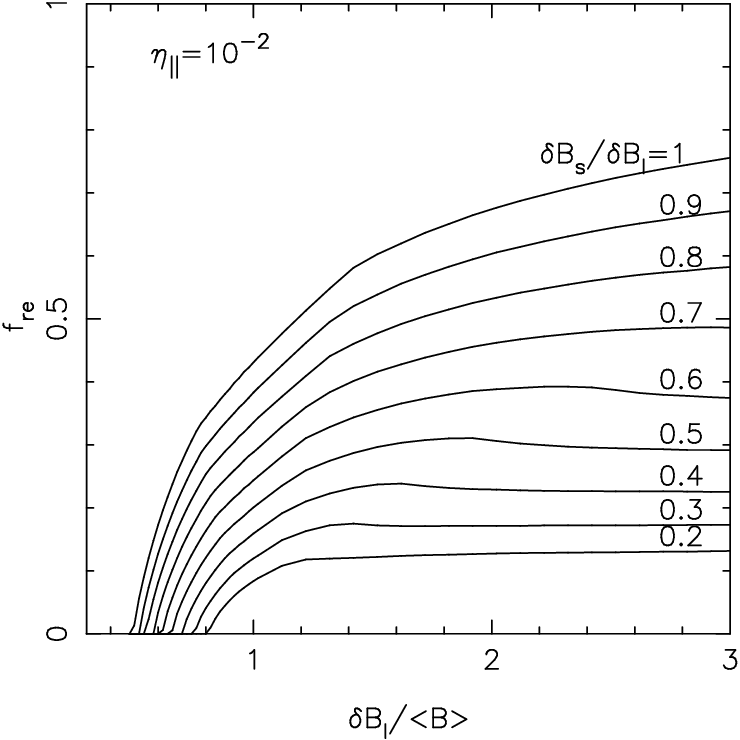}
\caption{The renormalization factor $f_{re}$ as a function of the random
large-scale magnetic field component, for different values of the
small-scale field component $\delta B_s$.}
\label{f4} 
\end{figure}

We shall assume an analogous magnetic field structure in the outer free
RSG wind region (Berezhko et al. \cite{bpv03}). The relevant instability
there is that of radiation pressure-driven winds (Lucy \& White
\cite{luw80};  for a more recent discussion, see Lucy \cite{lu84})
which we consider here in terms of the dynamical effect of the intense WR
star radiation field on the free RSG wind.

During the SNR shock propagation through the RSG wind material, the
variation of the random magnetic field component will lead in both phases
-- compressed shell and free RSG wind -- to a number of irregular regions
(spots) on the shock surface where the shock is quasi-parallel, whereas
the main fraction of the shock is quasi-perpendicular\footnote{Similar
regions with irregular shock normal angles relative to the magnetic field
will be produced by the clumpy nature of the SN ejecta -- the fast moving
knots -- which give the SNR shock a rather frayed appearance. For lack of
better knowledge we shall lump the two effects together here into a single
component $\delta B_l$.}.

The fraction of the shock surface $f_{re}$ where the shock is
quasi-parallel and where therefore efficient injection occurs,
depends on the ratios of the values of the magnetic field components
$\langle B\rangle$, $\delta B_l$ and $\delta B_s$. In order to estimate
the expected value of the renormalization factor $f_{re}$ for given values
of $\langle B\rangle$, $\delta B_l$ and $\delta B_s$, we first calculate the
injection rate $\eta(\theta_1)$ by performing the averaging procedure over
all possible directions of the small scale random field $\vec{B_s}$
according to the expression (\ref{eq9}), where in this case
$\vec{B_1}=\langle\vec{B}\rangle+\vec{B_l}$ and $\delta \vec{B}=\delta
\vec{B_s}$. Subsequently, we can calculate the value of the solid angle
$\Delta \Omega$ which includes those components of the large scale random
field $\delta \vec{B_l}$ whose directions lead to efficient injection,
that is $\eta(\theta_1)>\eta_{min}$. In the spirit of Fig.3, we then
assume that above $\eta_{min}$ injection rises nonlinearly as a result of
strong acceleration, if the typical scales of the high-injection flux
tubes exceed the size $l(p_{max}) \approx 0.1R_s$ of the CR precursor.  
This is expected to be the case at least in the compressed shell of Cas~A,
judging from the numerical results of Garcia-Segura et al.(\cite{garcia}).
In the outer free RSG region we have also assumed this to be true in the
model of Berezhko et al. (\cite{bpv03}). However, it is not clear how large
the radial scales of the radiation-induced instability are, even though
the scales of the frayed shock due to the fast moving ejecta knots should
be the same as in the shell. Therefore the efficient acceleration also in
the free RSG wind remains an assumption which we shall also make here.
Quantitatively, the radio and X-ray flux from the dense shell dominates
the overall nonthermal emission in these wavelength regions at the present
epoch (Berezhko et al. \cite{bpv03}), so that the contribution
from the free RSG wind is not a decisive factor in the interpretation of
the gamma-ray emission from Cas~A.

Since the gas number density in the shell is $N_g=10$~cm$^{-3}$ and the
magnetic field value is about $B_0=200$~$\mu$G, the Alfv\'en speed is
$c_a=130$~km/s which gives $\eta_{min}=1.4\times 10^{-5}(\delta B/B)^2_0$,
where $\delta B= \delta B_s$ and $B=\langle B\rangle +\delta B_l$. Because
the large-scale field component $\delta\vec{B_l}$ is randomly distributed,
the value of the renormalization factor is determined by the simple
expression
\begin{equation}
f_{re}=\Delta\Omega/(4\pi).  
\label{eq22}
\end{equation}
The value of the renormalization factor $f_{re}$ as a function of $\delta
B_l/\langle B\rangle$, calculated for different ratios of small-scale to
large-scale field components $\delta B_s/\delta B_l$, is shown in
Fig.~\ref{f4}. One can see that for $\delta B_l \sim \langle B\rangle $
the renormalization factor $f_{re}$ has values between 0.1 and 0.2 if the
small-scale random component has a relative amplitude $\delta B_s/\delta
B_l$ between 0.2 and 0.4. Such relative amplitudes appear quite realistic.

We conclude that the value $f_{re}=0.15$, which yields a satisfactory
description of all relevant properties of the emission generated in Cas~A
by accelerated CRs (Berezhko et al. \cite{bpv03}), is indeed consistent
with the expected structural properties of the circumstellar magnetic
field.

\section{Summary}

The injection rate of suprathermal particles into the shock acceleration
process depends strongly on the shock obliquity and diminishes as the
angle between the ambient field and the shock normal increases.  For the
ideal case of a SN explosion into a uniform interstellar medium with a
uniform magnetic field $\vec{B}$, efficient particle injection, leading to
the conversion of a significant fraction of the kinetic energy at a shock
surface element, only occurs in relatively small regions near the "poles",
reducing the overall CR production. The sizes of these regions depend
strongly on the random background field and on the Alfv\'en wave
turbulence generated upstream of the shock due to the CR streaming
instability.

For SN~1006 which appears to approximate this ideal case, efficient CR
production is expected to arise within two polar regions, where the SN
shock is quasi-parallel. The relative size of these regions depends
decisively on the amplitude of the random background field component
$\delta B$: it changes from a fraction of about 0.07 to one of 0.7 of the
shock surface when the interstellar random field amplitude varies from
$\delta B/B=0$ to 1, respectively.  It is argued that the nonlinear
backreaction of wave production on the injection rate leads to a definite
size of the injection fraction.

It is also argued that the actual spatially integrated nuclear gamma-ray
emission from such objects can be obtained through a renormalization of
the spherically symmetric result by the same factor.

For SN~1006, the calculated size of the efficient CR production regions
which amounts to about 20\% of the shock surface corresponds very well to
the observed sizes of the bright X-ray synchrotron emission regions (e.g.
Koyama et al. \cite{koyama};
Allen et al. \cite{allen}). This implies a renormalization factor of 0.2.

In the case of Tycho's SNR, with its unexpectedly nonuniform ambient
interstellar medium, we can not expect such a simple X-ray morphology.  
However, at least the topology of the field should be the same. Therefore
the renormalization factor $f_{re}$ has been calculated in an analogous
manner. Although the relevant parameters differ from those of SN~1006, we
obtain the similar value $f_{re}=0.18$.

The total size of the regions where efficient injection of suprathermal
particles occurs and the relevance of GCR reacceleration are expected to
increase during SNR evolution. This leads to a steepening of the resultant
energy spectrum of CRs produced in SNRs compared with the spherically
symmetric case.

In the case of a circumstellar medium which is strongly perturbed by the
mass loss of the progenitor star, like it exists for Cas~A, efficient
injection presumably takes place in a number of randomly distributed
portions of the shock surface with quasi-parallel magnetic field. The
fraction of the shock surface covered by these spots depends on the
relative strength of the mean magnetic field, which is assumed to be
tangential to the shock surface, and upon the amplitude ratio of the
small-scale and the large-scale scale random field components.  It is
larger for lower mean field values and for higher small-scale random
fields. It was demonstrated that quite reasonable magnetic field
parameters are consistent with efficient injection on about 15\% of the
shock surface, the fraction which is required in order to reproduce the
observed properties of Cas~A (Berezhko et al. \cite{bpv03}).

\begin{acknowledgements}
This work has been supported in
part by the Russian Foundation for Basic Research (grants 03-02-16325,
99-02-16325). EGB and LTK acknowledge the hospitality of the
Max-Planck-Institut f\"ur Kernphysik, where part of this work was
carried out. LTK also acknowledges the receipt of JSPS Research
Fellowship.
\end{acknowledgements}

\end{document}